\documentclass{revtex4}
\usepackage{epsf}
\usepackage{amsmath}
\usepackage{amssymb}

\renewcommand{\v}[1]{{\mathbf #1}}

\newcommand{\vx}{\v{x}}
\newcommand{\vC}{\v{R}}

\newcommand{\bea}{\begin{eqnarray}}
\newcommand{\eea}{\end{eqnarray}}
\newcommand{\be}{\begin{equation}}
\newcommand{\ee}{\end{equation}}
\newcommand{\Det}{\mbox{Det\,}}
\newcommand{\bc}[2]{\left(\begin{array}{c} #1 \\ #2\end{array}\right)}

\renewcommand{\r}{r}
\renewcommand{\c}{c}
\newcommand{\rh}{\rho}

\begin{document}

\title{Multiscale Complexity of Correlated Gaussians}
\author{Richard Metzler}
\affiliation{New England Complex Systems Institute,
 24 Mt. Auburn St., Cambridge, MA 02138, USA}
\affiliation{Department of Physics, Massachusetts Institute of Technology,
 Cambridge, MA 02139, USA}
\author{Yaneer Bar-Yam}
\affiliation{New England Complex Systems Institute,
 24 Mt. Auburn St., Cambridge, MA 02138, USA}

\begin{abstract}
We apply a recently developed measure of multiscale complexity to the
Gaussian model consisting of continuous spins with bilinear interactions
for a variety of interaction matrix structures.
We find two universal behaviors of the complexity profile.
For systems with variables that are not frustrated, an exponential decay 
of multiscale complexity in the disordered regime shows the presence 
of small-scale fluctuations, and a logarithmically diverging profile of fixed
shape near the critical point describes the spectrum of 
collective modes. For frustrated variables, oscillations in complexity
indicate the presence of global or local constraints. These observations
show that the multiscale complexity may be a useful tool for interpreting the 
underlying structure of systems for which pair correlations can be measured.
\end{abstract}
\maketitle
With the rising interest in complex systems comes a desire to quantify how
complex a given system is. Tools that may be used come
from information theory \cite{Shannon,Han:Mathematics,Goldie:Communication}
and statistical physics \cite{Reif:Fundamentals,Balian:Microphysics}. They
address questions about how long a complete microscopic description of a system
must be, given its constituents and constraints. 
However, the microscopic
information (the entropy) of a system does not correspond well with
intuitive concepts of complexity---a system with maximal entropy is merely
random, whereas a system with minimal entropy is strongly ordered;  
systems commonly considered complex, on the other hand, have rich internal
structure, i.e., constraints and correlations. 

This problem may be resolved by the recent approach of  
considering the complexity on different scales 
\cite{Bar-Yam:Dynamics, Bar-Yam:Multiscale, Speranta:Multiscale}:
random systems have high information content on small scales; however,
microscopic degrees of freedom average out over larger scales, leading to a
satisfying description in terms of a small number of 
variables such as volume, pressure and
temperature. Highly constrained systems, on the other hand, have roughly the
same information content on all length scales. E.g., one can describe a
system of many strongly bound particles by the location, orientation, 
and movement of the
center of gravity; given the (constant) positions of the particles relative
to each other, these few variables then determine everything there is to
know down to microscopic scales.
In contrast, truly complex systems have intermediate levels of
organization. For example, a human being has interesting behaviors on a
macroscopic level (the length scale of meters), which arises from the
level of organs (centimeters), which are composed of cells (micrometers),
which consist of biomolecules (nanometers).

A recently developed formalism \cite{Bar-Yam:Multiscale} allows us to
calculate the complexity on different levels of observation from the
underlying probability distribution of the degrees of freedom of the
system, which implicitly contains the interactions and constraints of the
system. The purpose of this paper is to apply this formalism to Gaussian
probability distributions with various structures of the covariance
matrix, yielding insight into the workings of the formalism, as well as
order-disorder transitions in systems with bilinear potentials.

The Gaussian distribution is also used in the  statistical analysis of
biological and social systems
\cite{Hartnett:Intro,Guttman:Linear}. 
The entries of the correlation matrix are often 
extracted from experimental data. The beauty of Gaussians is that their
probability distribution is
completely specified by two-point correlations; however, higher
correlations exist. As the paper will show, the multiscale
formalism offers an opportunity of detecting subsets of variables that form
coupled 
functional units, of testing for global or local constraints that manifest
themselves in the correlation matrix, and for
characterizing collective behaviors of systems.

The paper is organized as follows: 
Section \ref{SEC-Gaussian} reviews properties of correlated Gaussian
variables. Sec. \ref{SEC-Physical} reviews the mathematical representation
of physical degrees of freedom in the form of Gaussians. 
Sec. \ref{SEC-Multiscale}
gives an overview of the multiscale complexity formalism and previous
results. Section \ref{SEC-Examples} then shows how the formalism applies to
a variety of different interaction matrices. 
Section \ref{SEC-Results} summarizes and interprets the results. 

\section{Gaussian variables}
\label{SEC-Gaussian}
Due to the pervasive power of the central limit theorem that gives rise to
Gaussian distributions and due to their mathematical convenience,
Gaussians are the default assumption for probability distributions under
many circumstances. This paper discusses sets of $n$ Gaussian variables,
 labeled $x_i$, with $i\in \{1, n\}$. Each has a  mean of 0 and
a variance of $\sigma_i^2$: $\langle x_i\rangle=0$, $\langle
x_i^2\rangle=\sigma_i^2$. The
cross-correlations are given by the elements of the covariance matrix $\vC$:
$\langle x_i x_j \rangle = R_{ij}$. The joint probability distribution is
then given by 
\be
P(\vx)= \frac{1}{\sqrt{ (2\pi)^n \Det \vC}} \exp\left( -\frac{1}{2}
  \vx^{T}\cdot\vC^{-1}\cdot\vx  \right). \label{MULT-jointdist}
\ee
To calculate how much information is contained in a Gaussian variable, we
use Shannon's information theory \cite{Shannon}; 
however, for convenience, we use
natural units (base $e$) rather than bits; i.e., for one degree of 
freedom, the definition is
\be
I_x=- \int \ln(p(x))  p(x) dx.
\ee
for a single Gaussian of variance $\sigma^2$ this results in 
$I^G_x= (\ln(2\pi) +1)/2 + \ln(\sigma)$. 
For the joint distribution in Eq.(\ref{MULT-jointdist}), 
one can obtain
\be
I_{\vx} =  (n [\ln(2\pi) +1] + \ln(\Det\vC) )/2. \label{MULTI-DET}
\ee 
For the trivial case of $n$
uncorrelated Gaussians of variance $\sigma^2$, one has  
$\ln(\Det\vC)=2 n \ln{\sigma}$, so that the information is
just $n$ times that of a single variable.

\section{Physical interpretation}
\label{SEC-Physical}
There is a close formal analogy between Eq. (\ref{MULT-jointdist}) and the
canonical distribution for continuous degrees of freedom with bilinear
interactions $H= -\frac{1}{2}\sum_{ij} J_{ij}x_i x_j$: 
\be
P(\vx) = \frac{\exp\left(\frac{\beta}{2} \sum_{i, j} J_{ij} x_i x_j \right)}{Z},
\ee
where the partition function $Z$ is the integral of the numerator over state
space. We will consider two quite distinct interpretations of such a
system: the first is an individual 
particle in a harmonic potential with $n$ degrees of freedom, with $J_{ii}$
giving the potential along the coordinate axes, and $J_{ij}$
determining the shape of the potential along the diagonals.

The second interpretation considers each degree of freedom as a spin that
interacts with other spins. In the context of models for 
magnetic systems, this is called the ``Gaussian model'' \cite{Parisi} and 
has been used as an approximation 
to binary Ising spins. In the Gaussian model, it
is usually assumed that spins in the absence of interactions
follow a Gaussian distribution of variance 1 (independent of temperature), 
while off-diagonal interactions are weighted with a factor of $\beta$, and
there is no explicit self-interaction. 

In this paper, for convenience, we 
consider the self-interaction, which serves to keep spins bounded, as a
part of the interaction matrix. The relevant control parameter then becomes
the ratio of the self-interaction to the interaction with other spins,
rather than the temperature.
We can  then write the covariance matrix as  $\v{R}= (-\beta \v{J})^{-1}$. 
Changing $\beta$ or changing the magnitude of both
self-interaction $J_{ii}$ and off-diagonal interaction $J_{ij}$
 only results in a rescaling of
the variables: 
$\beta \sum x_i J_{ij} x_j = \sum x'_i J_{ij} x'_j$ with 
$x'_i= \sqrt{\beta}x_i$. This is a reflection
of the equipartition theorem, which states that each degree of freedom
corresponding to a quadratic term in the energy carries a mean energy of
$k_B T/2$.  As we show in Sec. \ref{SEC-Multiscale}, such a rescaling 
only adds an
additive term to the microscopic entropy, and does not change  
the terms in the multiscale complexity for scales larger than one. 
We can therefore set 
$\beta=1$ and the self-interaction $J_{ii}=-1$ unless otherwise
stated, whereas the interaction with neighbors is proportional to a 
parameter $a$ which we vary to control the system's behavior.

In contrast to the Ising model, the Gaussian model does not have a 
well-defined ordered phase. 
Interpreting the system as a particle in a harmonic potential, 
we find that  $\langle x_i^2 \rangle$ is finite 
(i.e., the system is in  the disordered phase) 
if all eigenvalues of $\v{J}$ are positive---otherwise there is
at least one degree of freedom with infinite negative energy, i.e., a harmonic
potential of the form $E(x)= - \lambda x^2$, $\lambda >0$. 
To model phase transitions more realistically in this
framework, one would have to introduce quartic terms \cite{Landau};
however, in this paper we will restrict ourselves to the pure Gaussian model. 
Interestingly, the transition to unbounded spins can happen in systems of
any size, and be due to the interaction of just two spins. 

The harmonic potential interpretation also offers a geometric argument
why the entropy is related to the determinant of the covariance matrix
(Eq. (\ref{MULTI-DET})): 
the coordinate system can be chosen such that
the main axes of the equipotential ellipsoid coincide with the coordinate
axes (i.e., the interaction matrix is diagonalized). This is always
possible because the interaction matrix is symmetric.
  The quadratic degrees
of freedom then decouple; the mean square amplitude along each axis depends
on the corresponding eigenvalue of the interaction matrix, and the
information (which is a measure of the size of state space) is the sum of
the information for the individual degrees of freedom. Specifically, it is
related to  
the sum of the logarithms of the eigenvalues, or equivalently the logarithm
of the determinant. Some of the axes of the new coordinate system can be
interpreted as aggregate variables such as the magnetization (the sum of
all spins).

In the diagonalized form, the susceptibility to external fields at the
transition is easy to
calculate. In the generic case, the smallest eigenvalue $\lambda_1$ 
can be linearly expanded around $0$: $\lambda_1 =
K(a-a_c)$. Applying a field $h$ along the eigenvector of $\lambda_1$ then
shifts the minimum of the energy $H$ (i.e., induces an magnetization) to
$h/2(K(a-a_c))$ along this axis. The susceptibility is therefore
proportional to $(a-a_c)^{-\gamma}$ with $\gamma=1$, independent of the
underlying dimensionality of the system. Different exponents
can arise only if the linear expansion of $\lambda_1$ around 0 has a
vanishing first derivative, which is not the case for any of the cases we
study in the following. 

The link between the entropy (or equivalently, information) 
and $\ln \Det \vC$ has one mathematical difficulty: 
since $\Det \vC$ can go to 0, the
entropy can diverge to negative infinity. 
The determinant vanishes if the rank of $\vC$ is smaller than $n$, i.e.,
one or more columns can be expressed as linear combinations of others. In
our context, this corresponds to a variable that is completely specified by
a combination of others. 
 Generally, the determinant is a polynomial of
order $n$ in the coefficients of the matrix, and the order of the zero in
question determines how many redundant variables there are. The diverging
entropy is not a problem for physical systems, since no physical quantity
can be perfectly specified.

\section{Multiscale complexity formalism}
\label{SEC-Multiscale}
A suitable measure of multiscale complexity $C_n(k)$ should fulfill several
conditions. For the smallest scale, it should correspond to the microscopic
entropy, which is the information contained in the joint probability
distribution of all degrees of freedom. 
If the system is composed of  distinct subsets of $l$ variables that are 
coupled within the subset, but not coupled to other subsets, $C_n(k)$ 
should be 0 for $k>l$, and take non-vanishing values corresponding to the
number of degrees of freedom otherwise. Furthermore, the multiscale
complexity of a composite of independent subsystems should be the sum of
subsystem complexities.

It has been shown in \cite{Bar-Yam:Multiscale} that the following
definition uniquely fulfills these conditions:
\bea
C_n(k) &=&\sum_{j=0}^{k-1} (-1)^{k-j-1} \left ( \begin{array}{c} n-j-1 \\ k-j-1
  \end{array}\right) Q(n,j), \mbox{~~with} \label{MULTI-complex}\\ 
Q(n,k)&=& -\sum_{\{j_1,\dots,j_k\}} \int \prod_{i_{k+1}}^{ i_n}dx_i P(\vx
-\{x_j\}) \ln P(\vx- \{\vx_j\}) \label{MULTI-Q}.
\eea
Eq. (\ref{MULTI-Q}) is a sum over all subsets of $k$ variables of the
entropy of the system after the subset has been removed. Thus, $Q(n,0)$ is
the microscopic entropy, and $Q(n,n-1)$ is the sum of entropies 
for individual degrees of freedom.

It is sometimes convenient to discuss the incremental difference in complexity
between scales $D(k)=C(k+1)-C(k)$. While $C(k)$ represents the effective 
number of degrees of freedom of size $k$ or larger, $D(k)$ represents the 
effective number of degrees of freedom at scale $k$.

The quantity defined by Eq. (\ref{MULTI-complex}) has some additional,
rather surprising, properties. In particular, it can have oscillations and take
negative values even for discrete variables. 
These seemingly anomalous properties 
have been shown to reflect the structure of the system, and specifically
have been linked to the effect of global constraints
on local variables, or more generally the concept of strong emergence, 
in Ref. \cite{Bar-Yam:Theory}. 
We will explore the relevance of such behavior to the Gaussian model.

To simplify the discussion further, we will show first that rescaling the
variables (e.g., by choosing a different inverse temperature $\beta$, as
described in Sec. \ref{SEC-Physical}) adds a term 
to the  microscopic entropy $C_n(1)$
that only depends on $n$ and the scaling ratio, and does not
affect higher-order complexities ($k>1$).
When applied to Gaussians, the terms $Q(n,k)$ are sums over the
logarithm of determinants of submatrices of $\v{R}$, dropping combinations
of $k$ rows and corresponding columns. Rescaling the covariance
 matrix by a factor $f$
 yields an additive term of $l \ln f$ to the
log of the determinant of a submatrix of size $l\times l$. Since there are
$\tbinom{n}{j}$ such contributions for $Q(n,j)$, the difference between the
rescaled multiscale complexity $\hat{C}_n(k)$ and the original is
\bea 
\hat{C}_n(k) -C_n(k)&=& \sum_{j=0}^{k-1}(-1)^{k-j-1}\binom{n-j-1}{k-j-1}
\binom{n}{j} (n-j) \ln f \nonumber \\
&=& \sum_{j=0}^{k-1} (-1)^{k-j-1}\frac{(n-j-1)!}{(k-j-1)!(n-k)!}
\frac{n! (n-j)}{j!(n-j)!} \ln f \nonumber \\
&=& \binom{n}{k} k \ln f \sum_{j=0}^{k-1} (-1)^{k-j-1}\binom{k-1}{j} = 
\left \{ \begin{array}{ll}  0 & \mbox{~~for~~}k >1 \\ n \ln f &
    \mbox{~~for~~} k=1 \end{array}\right. .
\eea
The identity in the last line can be derived by expanding
$(1+(-1))^{k-1}$ using binomial coefficients.


\section{Application to Specific Models}
\label{SEC-Examples}

\subsection{Three interacting spins}
\label{SEC-threespin}
We start by applying the formalism to
a minimal case of 3 interacting spins. The three-spin case 
captures a number of features that will be characteristic of larger spin 
systems. The complexity profile can be solved analytically giving:
\bea
C_3(1)&=& [3(\ln(2\pi)+1)-2 \ln(1+a) - \ln(1-2a)]/2;\nonumber \\
C_3(2)&=& -\ln(1+a) -\ln(1-2a)/2; \nonumber \\
C_3(3)&=& [- \ln(1-2a) + \ln(1+a) + 3\ln(1-a)]/2.
\label{MULTI-threespin}
\eea
\begin{figure}
\epsfxsize= 0.7 \columnwidth
  \epsffile{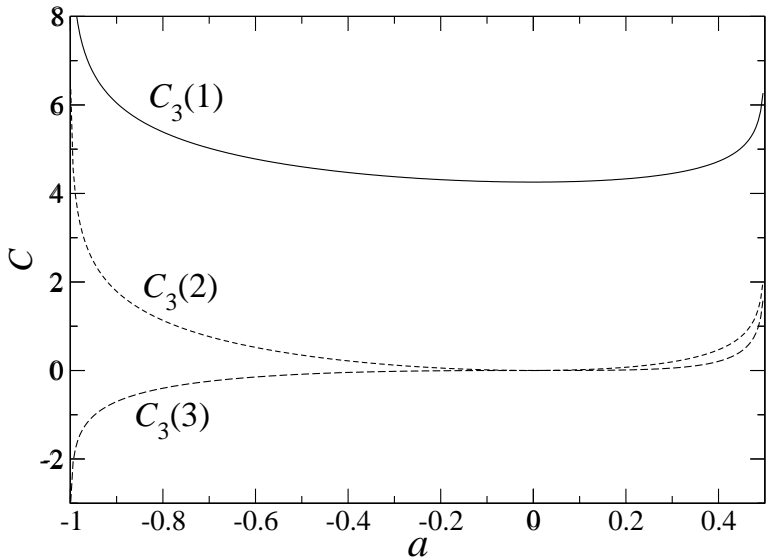}
   \caption{Multiscale complexities for the 3-spin chain. While all
     complexities are positive for $a>0$, $C_3(3)$ is negative for $a<0$.}
  \label{MULTI-FrustSpin}
\end{figure} 
As Fig. \ref{MULTI-FrustSpin} shows, complexities diverge at $a=1/2$ and
$a=-1$. These are the ferromagnetic and antiferromagnetic transitions to 
unbounded variables. As this case illustrates, unlike conventional binary
 spin systems, transitions occur even for finite numbers of spins. This result 
 can be understood by recognizing that conventional Gaussian variables 
 themselves can be thought of as arising as aggregates of many microscopic 
 bounded variables through the central limit theorem. 
 Negative values of  $C_3(3)$ occur for $a<0$. This is  expected for systems 
 with frustrated spins, due to the existence of a constraint on the three spins
 that does not affect any pair of spins. 

\subsection{Infinite range magnet}
\label{SEC-mfm}
For larger numbers of spins we consider first an infinite-range Gaussian
model, with an interaction matrix: 
\be
\v{J}=\left(\begin{array}{cccc}
-1 & a & \dots &a\\
a & -1 & a &\dots \\
a &\dots & \dots &a \\
a &\dots & a & -1
\end{array} \right), \label{MULTI-ferromat}
\ee
where $a$, the interaction between spins, 
is negative for antiferromagnets and positive for ferromagnets. Inverting
the negative of this matrix does not alter its structure; 
one obtains the covariance matrix
\be
\v{R}=\left(\begin{array}{cccc}
\r & \c & \dots &\c\\
\c & \r & c &\dots \\
\c &\dots & \dots &\c \\
\c &\dots &\c & \r
   \end{array} \right), \label{MULTI-ferrocorr}
\ee
with $\r$ and $\c$ given by 
\bea
\r&=& \frac{1-(n-2)a}{(1+a)(1-(n-1)a)}; \\
\c&=& \frac{a}{(1+a)(1-(n-1)a)}.
\eea
We define the correlation ratio $\rh=\c/\r$, in order to
separate the amplitude of the spins from the correlations between them:
\be
\rh = \frac{a}{1-(n-2)a}.
\ee
Since ferromagnetic ordering is easier to achieve with infinite-range
interactions than antiferromagnetic ordering, the transition where 
all spins collapse into one ($\rh$ goes to $1$) and the amplitude diverges
occurs at a small value of $a=1/(n-1)$. Antiferromagnetic ordering occurs
at $a=-1$, where 
$\rh$ takes the asymptotic value $-1/(n-1)$ and all
variables are maximally anticorrelated, taking values
corresponding to an $n-1$-dimensional hypertetrahedron.

The determinant of $\v{R}$ is 
\be
\Det\v{R_n}=r^n (1-\rh)^{n-1}(1+(n-1)\rh). \label{MULTI-DetMag}
\ee
The determinant has a zero of order $n-1$ at $\rh=1$, where all variables
are the same, 
and $n-1$ of them are redundant.
It also has a first-order zero at $\rh_c=-1/(n-1)$, where 
geometrical constraints can be used to eliminate one variable.

We can now calculate the scale-dependent complexity $C_n(k)$ following
Eq. (\ref{MULTI-complex}).
For our example, we can calculate this analytically for $k=1,2,3$ and
numerically for other values of $k$. Since the matrix
does not change its structure if one or more variables are removed, the
determinant has the same form as Eq. (\ref{MULTI-DetMag}) with a modified
number of variables. 

We then have 
\bea
Q(n,j)&=& \bc{n}{j} \left[ \frac{n-j}{2}(\ln(2\pi \r)+1  ) + \right. 
     \nonumber \\
&&\mbox{~~~~~}  
 \left. \frac{1}{2}\left((n-j-1)\ln(1-\rh) + \ln(1+(n-j-1)\rh)\right)\right],
\eea
which yields
\bea 
C_n(1)&=& \frac{n}{2}\left(\ln(2\pi) +1 + \ln(\r)\right) + 
\frac{n-1}{2}\ln(1-\rh) +
\frac{1}{2}\ln(1+(n-1)\rh); \label{MULTI-meanfieldC1} \\
C_n(2)&=& [-\ln(1-\rh) - (n-1) \ln(1+(n-1)\rh) + n\ln(1+(n-2)\rh)]/2; \\
C_n(3)&=& - \ln(1-\rh)/2 +(n-1)(n-2)\ln(1+(n-1)\rh)/4 - \nonumber \\
& & n(n-2)\ln(1+(n-2)\rh)/2 + n(n-1)\ln(1+(n-3)\rh)/4; 
\label{MULTI-meanfieldC3}
\eea
In terms of $a$, these complexities can be written as
\bea 
C_n(1)&=& \frac{n}{2}\left(\ln(2\pi) +1 + \ln(1-(n-1)a)\right) + 
\frac{n-1}{2}\ln(1-(n-1)a) +
\frac{1}{2}\ln(1+a); \label{MULTI-meanfieldaC1} \\
C_n(2)&=& [-\ln(1-(n-1)a) - (n-1) \ln(1+a)]/2; \\
C_n(3)&=& [- 2\ln(1-(n-1)a) +(n-1)(n-2)\ln(1+a) + n(n-1)\ln(1-a)]/4; 
\label{MULTI-meanfieldaC3}
\eea
For $n=3$ this reduces to the case of three interacting spins,
Eq. (\ref{MULTI-threespin}). All complexities for $k>1$ include a singular term
$-\ln(1-\rh)/2 = -\ln(1-(n-1)a)/2$ but are
independent of $r$, which is a consequence of the
invariance to multiplying all spins by the same factor, as described
above. $C_n(1)$, on the other hand, includes a term $\ln(2 \pi \r) +1$. 

\begin{figure}
\epsfxsize= 0.7 \columnwidth
  \epsffile{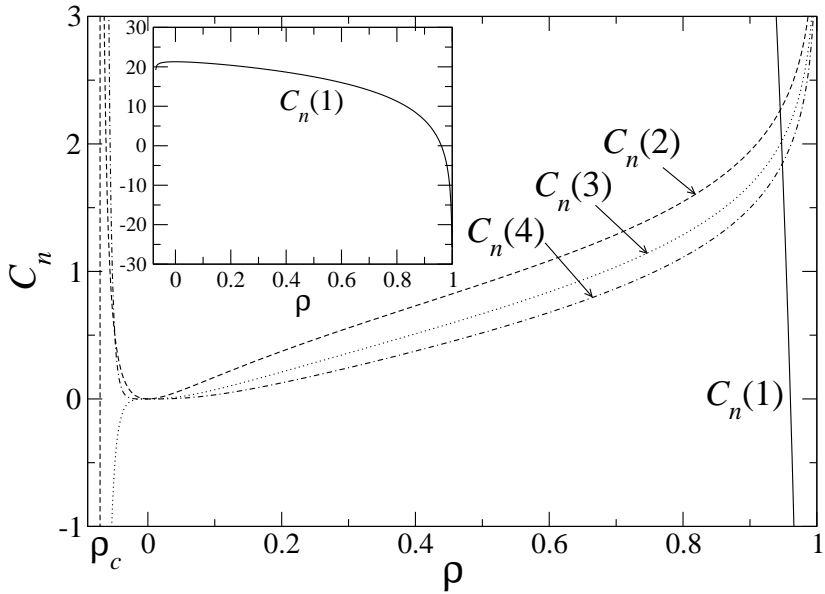}
   \caption{Complexities $C_n(1)$ to $C_n(4)$ for infinite-range
     interactions  
     (for $n=15$ and amplitude $\r=1$) following
     Eqs. (\ref{MULTI-meanfieldC1}-\ref{MULTI-meanfieldC3}). 
     The inset shows $C_n(1)$ for a
     different scaling of the $y$-axis.}
  \label{MULTI-P1}
\end{figure}

\begin{figure}
\epsfxsize= 0.7 \columnwidth
  \epsffile{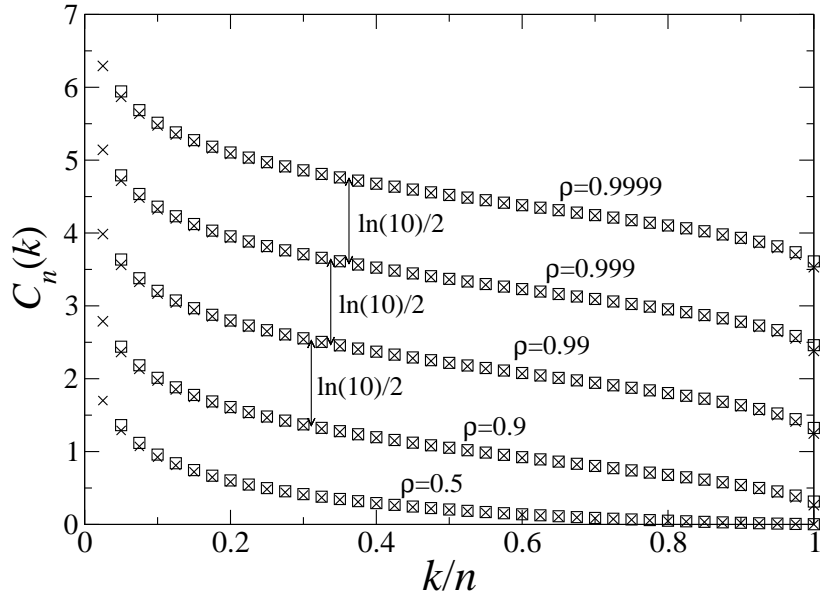}
   \caption{Complexities $C_n(k)$ for $n=40$ (squares) and $n=80$ (crosses)
     for the infinite-range ferromagnet, displayed as a function of $k/n$,
     for various values of $\rh$. Arrows indicate
     the shift predicted by Eq. (\ref{MULTI-meanfieldC3})}
  \label{MULTI-P2}
\end{figure}

We find two characteristic behaviors for ferromagnetic and 
antiferromagnetic interactions.
In the ferromagnetic case ($a>0$, $\rho>0$) the multiscale complexity is
positive, and decreases monotonically with increasing $k$.
The curves of $C_n(k)$ plotted as a function of $k/n$ approximately 
collapse to one curve for 
each value of $\rh>0$, so that the complexity can be written as a scaling
function $C(k/n, \rh)$. Near $\rh=1$, this function takes a
universal shape, so that 
$C(k/n,\rh)\approx C_l(k/n)+ E(\rh)$. The singular term in all complexities
pointed out above, $-\ln(1-\rh)/2$, 
can be identified with $E(\rh)$. As seen in Fig. \ref{MULTI-P2},
reducing $1-\rh=\rh_c-\rh$ [or equivalently $a_c-a$, as seen in
Eq. (\ref{MULTI-meanfieldaC1}-\ref{MULTI-meanfieldaC3})]
 by a factor of 10 increases
$C_n(k/n)$ by $\ln(10)/2$. 

The universal shape of $C(k/n,\rh)$ implies that $D(k)$, the spectrum of
excitations, is independent of $\rh$ near the transition. With increasing
coherence of the variables, the collective behavior at $C(n)$ increases
at the expense of the independence of the variables $C(1)$ without
affecting the spectrum of intermediate scale excitations.

\begin{figure}
\epsfxsize= 0.7 \columnwidth
  \epsffile{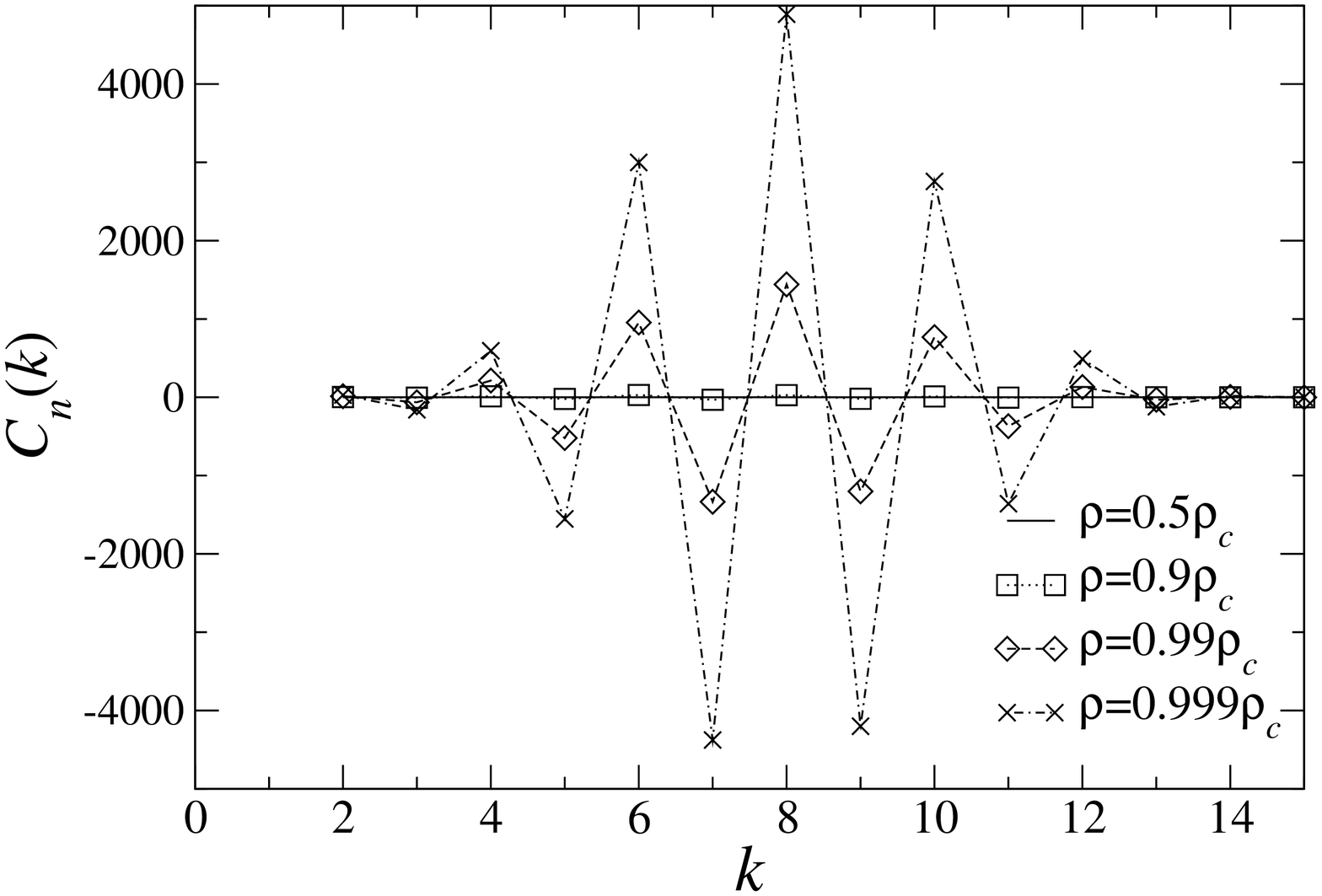}
  \caption{Complexities $C_n(k)$ for the infinite-range antiferromagnet
    with $n=15$, for different values of
    $\rho<0$, expressed as fractions of $\rho_{c}= -1/(n-1)$. 
    This is close to the point where one variable becomes
    redundant; the multiscale complexity shows oscillations of increasing 
    amplitude.}
  \label{MULTI-P3}
\end{figure}

In the antiferromagnetic case ($a<0, \rho<0$)
oscillations in $C_k(n)$ are found, as seen in Fig.  
\ref{MULTI-P3}. The amplitude of oscillations increases as $-f
\ln(1 + (n-1)\rho)-g$, where $f$ and $g$ increase rapidly with $n$ and 
are of order $10^5$ for $n=20$. Oscillatory behavior in 
the multiscale complexity has been
linked to global constraints \cite{Bar-Yam:Theory}; in this case, it is the
constraint $\sum x_i=0$ which is enforced by the interactions near the
transition point, and which can be used to eliminate one of the variables.

It should be pointed out that 
Eq. (\ref{MULTI-complex}) is susceptible to numerical inaccuracies,
and with insufficient precision one observes oscillations for $\rho>0$; 
however, these are numerical artifacts, as can be seen by increasing
the accuracy of calculations. The oscillations in the antiferromagnetic
regime are not artifacts.

\subsection{1-D spin chain}
\label{SEC-spinchain}
We consider a chain of $n$ spins with nearest-neighbor
interactions $a$ and self-interactions $-1$.  The interaction matrix is given by
\be
\v{J} = \left( \begin{array}{ccccc} 
-1 & a &0 &\dots & a \\
a &-1& a& 0& \dots \\
& & \dots & & \\
a &0 &\dots &a & -1 \end{array} \right )
\ee
For even $n$, it is not relevant whether interactions are
ferromagnetic or antiferromagnetic, since both cases can be mapped onto
each other by flipping every other spin. For odd $n$, the asymmetry with
respect to $a$ becomes less pronounced as $n$ increases: negative
complexities are only observed for $n\leq 5$, and for $n>15$ curves look
largely symmetric. The following observations assume even $n$.

For weak interactions, one finds correlations that decrease exponentially
with distance; specifically, the leading order term of $R_{ij}$ 
 is $a^{|i-j|}$. Due to the short-ranged, one-dimensional interactions,
 Ising spins with this interaction structure do not show an 
order/disorder phase transition. 
However, the continuous-spin system displays a transition from finite to
infinite variances for $a=\pm 1/2$, at which point the
nearest-neighbor interactions override the self-interaction.  

We find numerically that $C_n(k)$ is proportional to $n$ for small $k>1$: 
each additional spin takes a finite amount of
information to describe. This is consistent with the short-range
interactions of the model. 

\begin{figure}
\epsfxsize= 0.7 \columnwidth
  \epsffile{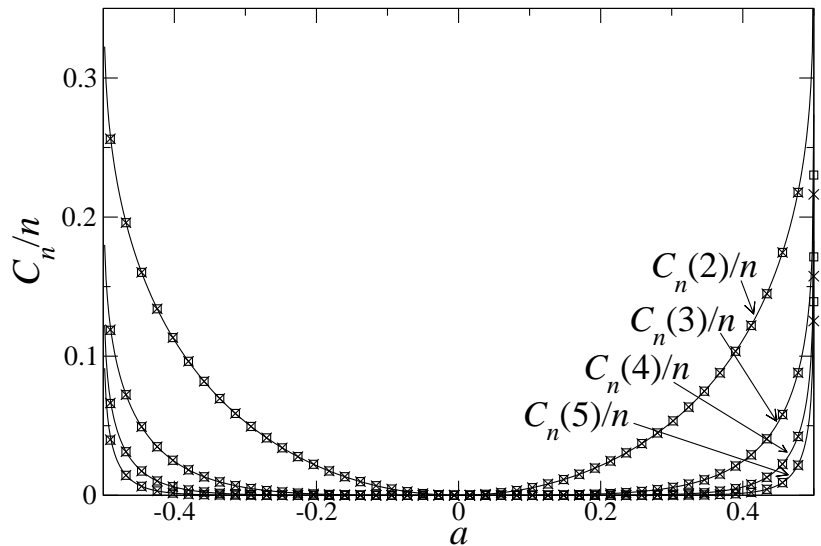}
   \caption{Multiscale complexities for the multi-spin chain, normalized by
     the number of spins $n$, for $n=16$ (lines), $n=25$ (squares) and
     $n=32$ (crosses). }
  \label{MULTI-spinchain}
\end{figure}
\begin{figure}
\epsfxsize= 0.7 \columnwidth
  \epsffile{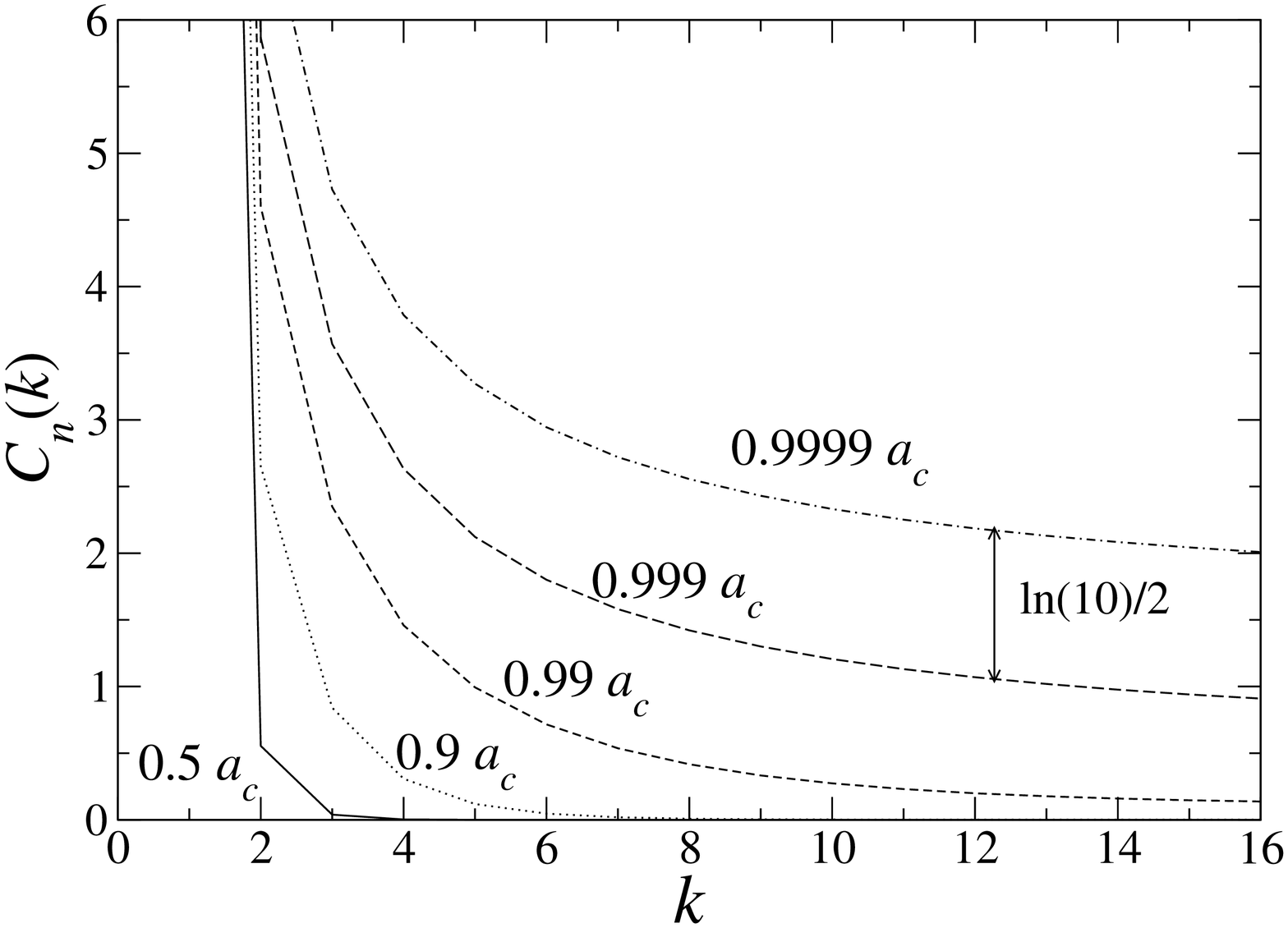}
   \caption{Multiscale complexity for the 1-D spin chain with 
     nearest-neighbor interactions, for different interaction strengths at 
     $n=16$.}
  \label{MULTI-spinchainprof}
\end{figure}

 
The complexity profile for all scales
(Fig. \ref{MULTI-spinchainprof}) behaves similarly
to the infinite range ferromagnet (Fig. \ref{MULTI-P2}): as $a$
approaches $a_c$, the complexity profile has a universal shape, $C_n^0(k)$,
that is monotonically decreasing and spans all scales. The divergence as a
function of $a-a_c$ follows $C_n(k,a) = C_n^0(k) - \ln(a_c-a)$.  

\subsection{Square lattice}
For a two dimensional example, we arrange spins on
a $L\times L$ square lattice with periodic boundary
conditions and nearest-neighbor
interactions of strength $a$. The transition point to unbounded spins is at
$a=\pm 1/4$. Intuitively, four interactions
of strength $1/4$ are sufficient to balance the self-interaction of strength 1.

As with the spin chain we find that the multiscale complexities
are roughly proportional to the number of spins in the system, and are almost
symmetric with respect to switching the sign of $a$. For even L the symmetry
is exact since a 
ferromagnetic square lattice can be transformed into an antiferromagnetic one
by flipping spins in a checkerboard pattern. 
The symmetry is not exact for odd $L$ and the deviation is largest for
small values of $L$.
For different $L$ the values of $C(k)/N$ do not collapse as well,
indicating a more significant deviation from extensivity than in the
one-dimensional case.

\begin{figure}
\epsfxsize= 0.7 \columnwidth
  \epsffile{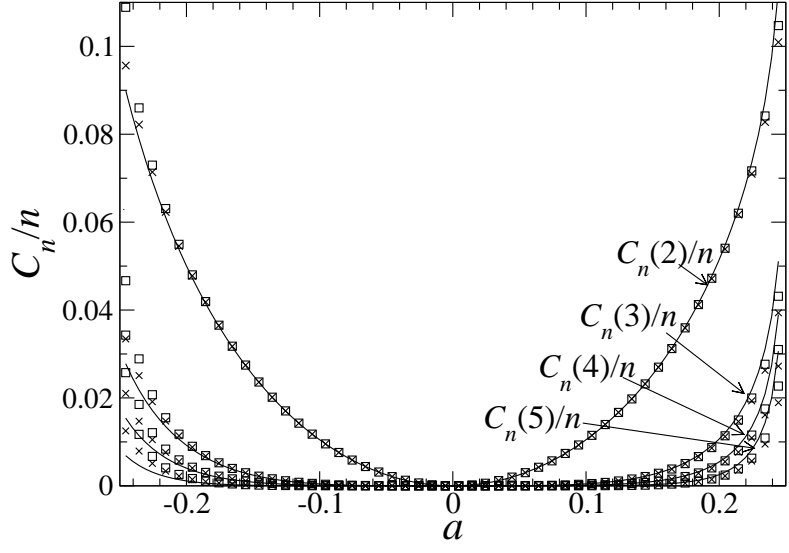}
   \caption{Multiscale complexities for the square lattice with 
     nearest-neighbor interactions, normalized by
     the number of spins $n=L\times L$, for $L=5$ (lines), $L=6$ (squares)
     and $L=7$ (crosses). }
  \label{MULTI-spinsheet}
\end{figure}
\begin{figure}
\epsfxsize= 0.7 \columnwidth
  \epsffile{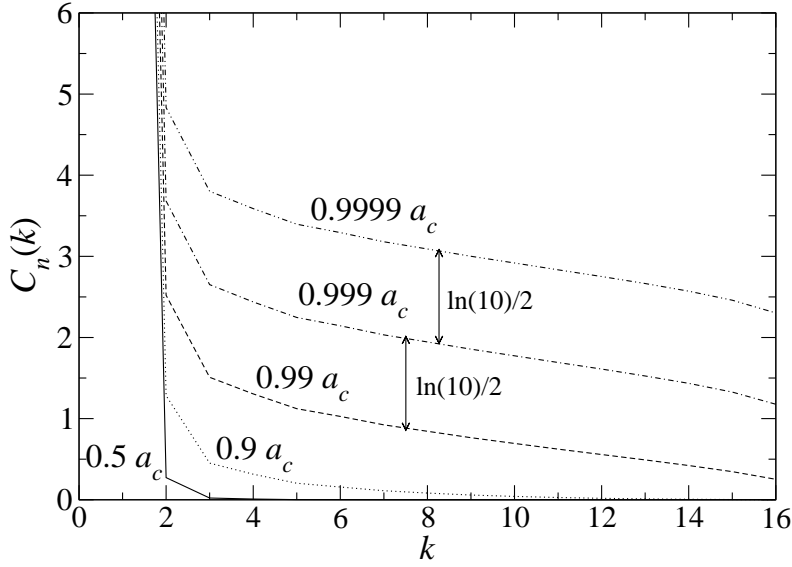}
   \caption{Multiscale complexity for the square lattice with 
     nearest-neighbor interactions, for different interaction strengths at 
     $L=4$.  }
  \label{MULTI-spinsheetprof}
\end{figure}

 
The complexity profile shows a similar behavior to the infinite-range 
ferromagnet: 
as $a$ approaches $a_c$, a complexity profile with universal shape emerges 
that spans all scales, and is monotonically
decreasing, as seen in Fig. \ref{MULTI-spinsheetprof}. A logarithmic
divergence as a function of $a-a_c$ is found as before. The shape of the   
complexity profile in the limit is different from the previous scenarios;
the cases will be compared below.

\subsection{Block matrix}
We generalize the infinite range Gaussian model by dividing the
spins into groups that interact strongly within the group, but weakly with
members of other groups, yielding the following interaction matrix (we
explicitly label the self-interaction as $-d$ in this case):
\be
\v{J} = -\left( \begin{array}{ccccccccc} 
d & \dots &  a &    &    &  &        &  &    \\
a  & d     &  a &       &  b  &  &        & b &      \\
a  & \dots & d &   &    &  &        &  &     \\
   &       &    &   d & \dots &  a & &    &   \\ 
   &   b   &    &   a  & d     &  a &   &  b  &    \\
   &       &    &   a  & \dots & d &  &    &   \\
    &       &    &  &       &    &   d & \dots &  a \\
 &   b   &    & &   b   &    &  a  &d      &  a \\
 &       &    &   &       &    & a  & \dots & d
 \end{array} \right ),
\label{Block-interact}
\ee
with $m$ blocks of $n$ spins each. This model is important in the study of
complex systems since it has a multi-level structure, i.e. modularity,
considered to be a universal property of complex systems 
\cite{Simon:Architecture,Bar-Yam:Dynamics}.

The determinant of this matrix is 
\be
\Det(-\v{J}) = (d-a)^{m(n-1)}(d+(n-1)a-nb)^{m-1}(d+(n-1)a+n(m-1)b), 
\label{Block-det}
\ee
which enables us to find the permissible ranges of $d$, $a$, and $b$:
for bounded spins to exist, $d$, which is the negative
self-interaction, has to be positive.
 Positive values of $a$ and $b$ imply 
antiferromagnetic
interactions, negative 
ferromagnetic. The boundaries of the allowed 
region are given by the zeros of Eq. (\ref{Block-det}):
\bea
a/d &\leq& 1; \label{Block-boundary1}\\ 
b/d &\leq&(1+(n-1)a/d)/n;\label{Block-boundary2} \\
b/d &\geq&-(1+(n-1)a/d)/(n (m-1)).\label{Block-boundary3}
\eea

\begin{figure}
\epsfxsize= 0.7 \columnwidth
  \epsffile{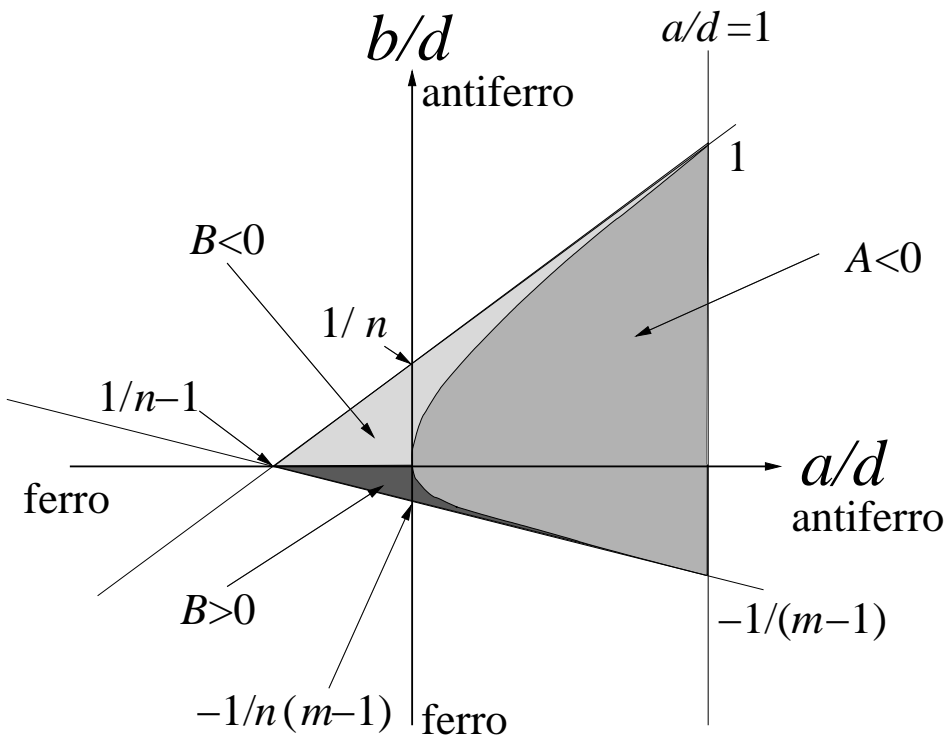}
   \caption{Phase diagram for block matrix with $m$ groups of $n$ spins
     each. Relevant variables are the ratios between near/far neighbor
     interactions and self-interaction. Shading indicates the sign of
     covariances $A$ and $B$.}
  \label{MULTI-blockphase}
\end{figure}

The inverse of this matrix has the same structure as Eq. 
(\ref{Block-interact}); if we label 
the coefficients of
$-\v{J}^{-1}$ as $A$, $B$, and $D$, respectively, we obtain
\bea
D &=& K\times  [ d^2 + (m-2)(n-2)n ab -(m-1)(n-1)n b^2 +\nonumber \\
 &&  (n-1)(n-2)a^2+( (m-2)nb + (2n-3)a)d ]; \nonumber \\
A &=& K\times\left(- ad  -(n-1)a^2 + (m-2)n ab + (m-1)n b^2 \right);
\nonumber \\
B &=& -K\times (d-a)b, \label{MULTI-blockinverse}
\eea
where 
\be
K = \left[ (d-a) (d+(n-1)a - n b)(1+(n-1)a+ (m-1)nb)\right]^{-1}.
\ee
Since the inverse of the negative interaction matrix is the covariance
matrix, this allows us to determine under what circumstances correlations
to near/ far neighbors are positive or negative. $A$ has a zero for
\be
b_{1,2}= \frac{1}{2(m-1)n}\left ((m-2)na \pm \sqrt{m^2n^2a^2 +
    4(m-1)n(d-a)a}\right); \label{Block-root}
\ee
it is negative between the branches of the root, and positive outside.
$B$ is negative for $b<0$, and positive for $b>0$. Combining the
results from Eqs. (\ref{Block-boundary1}-\ref{Block-boundary2}) and
(\ref{Block-root}), the phase diagram shown in Fig. \ref{MULTI-blockphase}
 emerges. Interestingly, long-range interactions do not
 stabilize the system: at any $b\neq 0$, the system has a smaller
 range of stability with respect to $a$ than for $b=0$.

We now calculate the multiscale complexities. $C_{nm}(1)$ 
can be calculated from the determinant Eq. (\ref{Block-det}) with entries
$D$, $A$, and $B$ from Eq. (\ref{MULTI-blockinverse}):
\bea
C_{nm}(1) &=& \frac{1}{2}\left\{ nm[\ln(2\pi) +1] + n (m-1) \ln(D-A)
  \right. 
 \nonumber \\
&& + (m-1)\ln(D + (n-1) A -nB) + \ln(D + (n-1)A + n(m-1)B).
\eea
Unfortunately, removing spins 
does not leave the structure of the covariance matrix unchanged. 
The determinant of the matrix with one spin removed is
\bea
\Det{C_{-1}} &=&  (D-A)^{m(n-1)-1} \left(D+ (n-1)A -nB\right )^{m-2}\times
\nonumber \\
&& \left( D^2 +(2n-3)AD +(n-1)(n-2)A^2 + \right. \nonumber \\
&& \left. n(m-2)(D+(n-2)A)B - n(n-1)(m-1)B^2\right).
\eea
From this expression, the corresponding complexity for scale 2 can be
found:
\bea
C_{nm}(2) &=& (1/2)[-m \ln(D-A) \nonumber \\
&& + (m-mn -1) \ln( D + (n-1) A- nB) \nonumber \\
&& - (mn -1) \ln (D + (n-1) A+ n (m-1) B) \nonumber \\
&&+ mn \ln (  D^2 +(2n-3)AD +(n-1)(n-2)A^2 + \nonumber \\ 
&& n(m-2)(D+(n-2)A)B - n(n-1)(m-1)B^2)]. \label{MULTI-blockC2}
\eea
The first term in Eq. (\ref{MULTI-blockC2}) indicates that the divergence
near $A=D$ is logarithmic in $m \ln(D-A)/2$; the additional factor of $m$,
compared to previous scenarios, indicates that there are $m$ distinct
units. 

\begin{figure}
\epsfxsize= 0.7 \columnwidth
  \epsffile{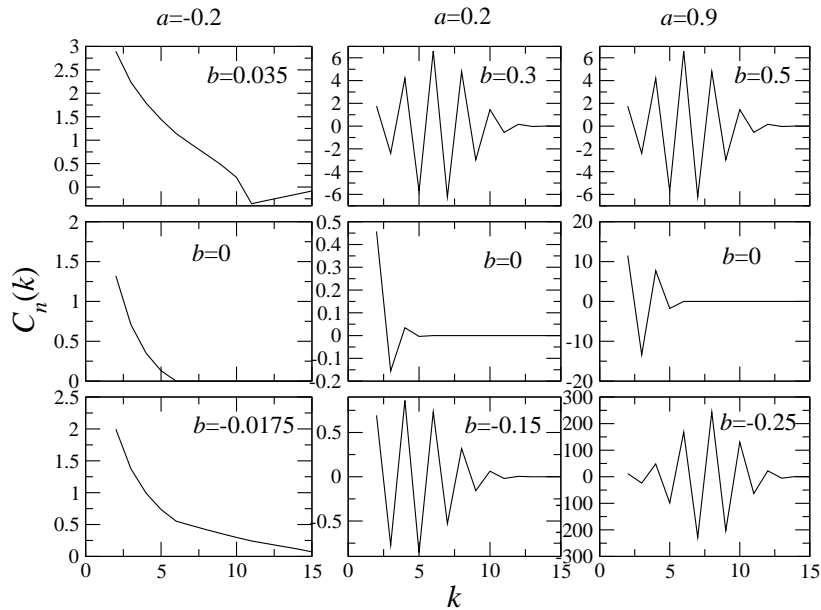}
   \caption{Multiscale complexities for block structure interactions with
     $n=5$, $m=3$, for different values of $a$ and $b$.}
  \label{MULTI-blockcomplexities}
\end{figure}

Expressions for higher $k$ do not give additional insight. We therefore 
turn to numerical results, which are shown in Fig. \ref{MULTI-blockcomplexities}.
Along the $b=0$ axis, the system consists of $m$ blocks of $n$ coupled
spins each. Complexities $C_{nm}(k)$ 
are different from zero for $k\leq n$ (positive for ferromagnetic
interactions, oscillating for antiferromagnetic), 
and equal to zero for $k>n$. This reflects the built-in property of the
multiscale formalism to identify non-interacting subsets of variables.

For $b\neq 0$, complexities at scales up to
$k=nm$ become non-zero. In the quadrant $a<0, b<0$ where all interactions
are ferromagnetic, all complexities are larger than zero. 
In the other quadrants,
one observes either negative complexities for higher scales or
oscillations similar to those found for the infinite-range antiferromagnet.

\subsection{Infinite-range spin glass}
Spin glasses (magnetic systems with random ferro- or antiferromagnetic
interactions) are a well-studied model of systems with
frustrated interactions, multiple degenerate ground states and other
interesting features \cite{Fischer:SpinGlasses,Chowdhury:SpinGlasses}.
The multiscale complexity of spin glasses is therefore of considerable
interest. We consider interaction matrices with diagonal elements
$J_{ii}=-1$ and
off-diagonal elements $J_{ij} = J_{ji} = a r_{ij}$, where $r_{ij}$ is a
Gaussian random variable of variance 1. To determine transition points for
any one realization of the quenched random interactions, we keep the set of
$r_{ij}$ constant and adjust the interaction strength $a$. 

The first quantity of interest is the critical $a$ at which the transition
from bounded to
unbounded variances occurs. This is related to the eigenvectors of $-J$,
which are necessarily all positive for bounded spins. The spectrum of
eigenvalues $\lambda$ of Gaussian matrices has been studied
\cite{Wigner1958}; if all entries (including the diagonal) are Gaussians of
mean 0 and variance $\sigma^2$, the matrix is symmetric, 
and $n$ is large, it takes the form 
\be
P(\lambda) = \left \{ \begin{array}{ll}
(2 \pi \sigma^2 n)^{-1}\sqrt{4 \sigma^2n - \lambda^2} & 
   \mbox{for} |\lambda| < 2 \sigma\sqrt {n} \\
0 & \mbox{else}
\end{array}  \right.,  \label{SPIN_circle}
\ee
i.e., a semi-circle of a width proportional
to $a\sqrt{n}$. For small $n$, the cutoff of this semi-circle becomes
blurred. 

Calculations show that replacing the random diagonal elements with
non-random elements of magnitude $d$ has two effects: $n$ is replaced by
$n-1$ in Eq. (\ref{SPIN_circle}), and the mean of the distribution is
shifted by $d$ (as can be shown analytically). Thus, the distribution of
$\lambda$ follows 
$P(\lambda) = \sqrt{4 \sigma^2 (n-1) - (\lambda-d)^2}/(2 \pi
\sigma^2 (n-1))$.

The critical interaction value $a_c$ is that for which the smallest
eigenvalue becomes $0$, for $d=1$. Neglecting the blurring of boundaries, 
we thus expect $|a_c| = d/\sqrt{4(n-1)}$, which is in good agreement with
calculations for large $n$. It should be pointed out that $a_c$ has significant
fluctuations for small $n$---the exact transition point differs 
for each realization of $\{ r_{ij}\}$. 
These result indicate a significant difference in the behavior of the transition
in the case of the spin glass and the infinite-range magnet with
uniform interactions. The scaling of
$a_c\propto n^{-1/2}$  lies between that of the infinite-range ferromagnet
($a_c\propto n^{-1}$) and the antiferromagnet ($a_c = 1$) described in
Sec. \ref{SEC-mfm}. 

\begin{figure}
\epsfxsize= 0.8 \columnwidth
  \epsffile{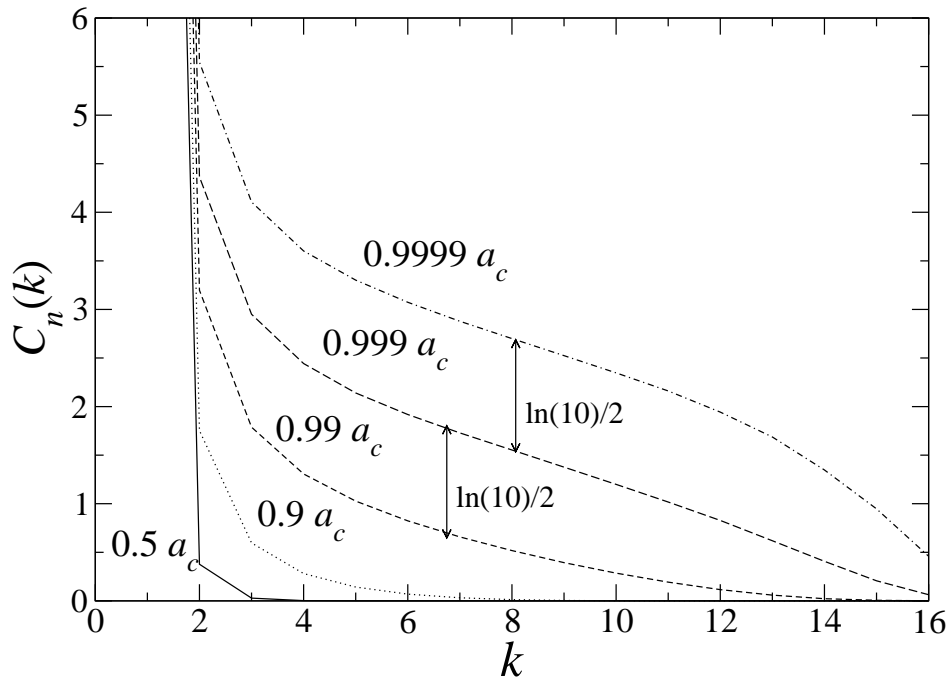}
   \caption{Complexity profile for long-range spin glass with
     $n=16$, for different values of $a/a_c$, averaged over 100
     realizations of randomness.}
  \label{MULTI-spinglass}
\end{figure}

Using these results, we explore the behavior of the
multiscale complexities $C_{n}(j)$. 
Near $a_c$, we obtain a monotonically decaying positive curve, similar to that
for the ferromagnetic mean-field magnet. For smaller
values of $a$, the curve is shifted to lower values of $C_n(j)$, but does
not become negative (see Fig. \ref{MULTI-spinglass}).

\begin{figure}
\epsfxsize= 0.8 \columnwidth
  \epsffile{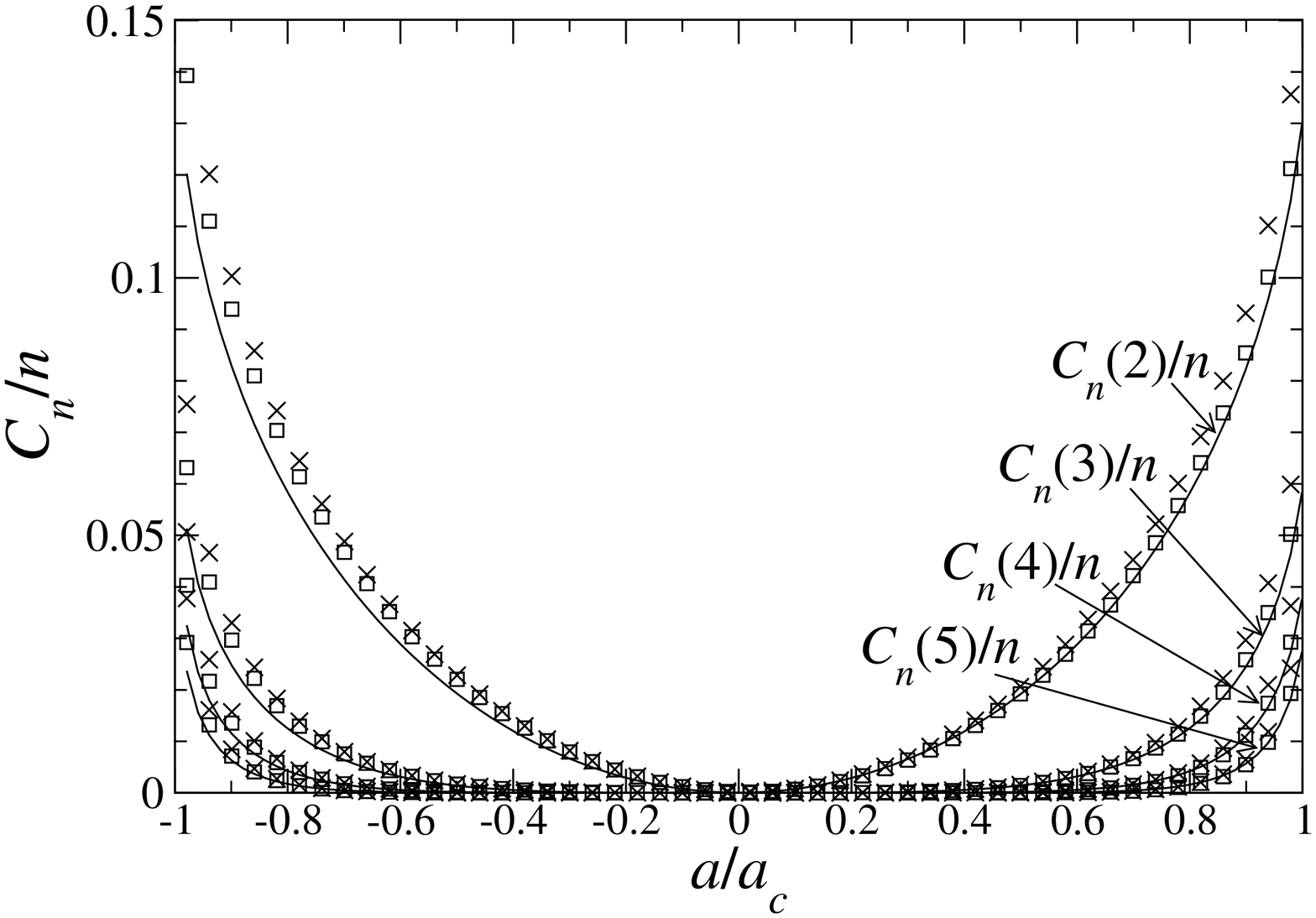}
   \caption{Multiscale complexities for the mean-field spin glass, 
     normalized by
     the number of spins $n$, for $n=40$ (lines), $n=30$ (squares) and
     $n=20$ (crosses), averaged over multiple realizations of randomness.}
  \label{MULTI-spinglasscoll}
\end{figure}

Interestingly, the curves of $C_n(k)$ for small $k$ follow similar
scaling behavior as those for ferromagnetic 
nearest-neighbor spin chains and lattices: when plotted as a function of
$a/a_c$, $C_n(k)/n$ roughly collapses onto one curve for different $n$, as
shown in Fig. \ref{MULTI-spinglasscoll}. 
The exact values of the curves
depend on the quenched variables, so that averaging over different
realizations becomes necessary; however, the 
finite-size effects decrease with increasing $n$.

It is significant that the random and partly frustrated
interactions do not lead to oscillations in the multiscale
complexity. Ferromagnetic modes become dominant and trigger the transition
before frustration has an impact. The study of spin-glasses using replica
symmetry breaking has previously found that the spin-glass order parameter
has a degeneracy of order $n$ in the ground state with spins having a
macroscopic ordering in the direction of one of these low energy
states. Such macroscopic ordering is indeed similar to the ordering found
in an infinite range model and is qualitatively different from a frustrated
infinite range model which constrains the macroscopic state to a
macroscopically degenerate subspace. Still, we note the different
scaling between frustrated and unfrustrated interactions described above.

\subsection{Triangular lattice}
\label{SEC-triangular}
\begin{figure}
\epsfxsize= 0.8 \columnwidth
  \epsffile{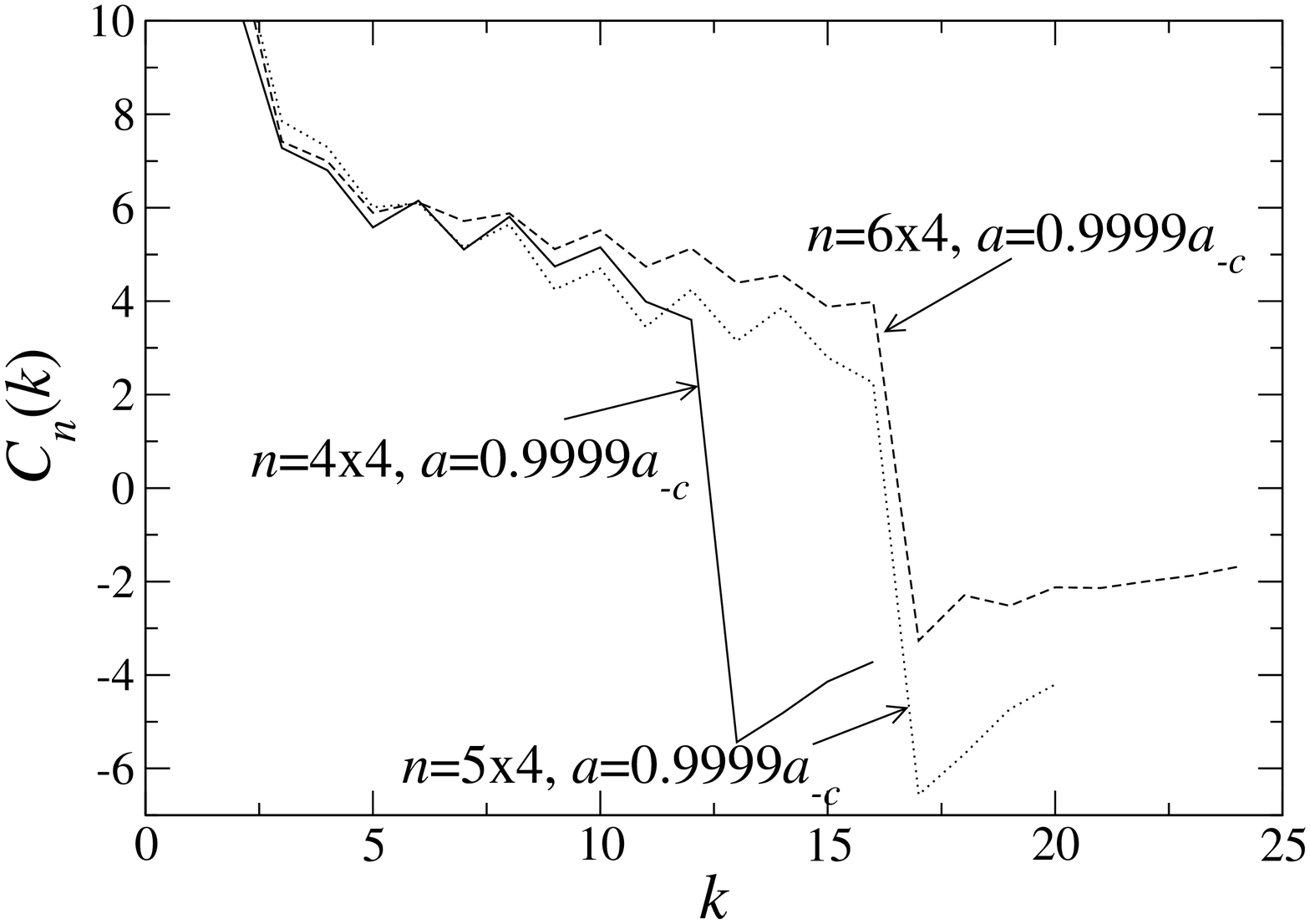}
   \caption{Complexity profile for the triangular lattice near the
     antiferromagnetic transition, for different system sizes.
     One sees signs of both frustration and coherent behavior.}
  \label{MULTI-triangularprof}
\end{figure}

Our final example is a two-dimensional lattice with frustration in the
anti-ferromagnetic state, the 
2-D triangular lattice. Again, we find two transitions points. In the
ferromagnetic regime, the transition is $a=1/6$ as expected for a
lattice of coordination number six. The complexity profile at the ferromagnetic
transition is qualitatively similar to that of the square-lattice
ferromagnet, and the multiscale complexity for small $k$ is extensive.
\begin{figure}
\epsfxsize= 0.5 \columnwidth
  \epsffile{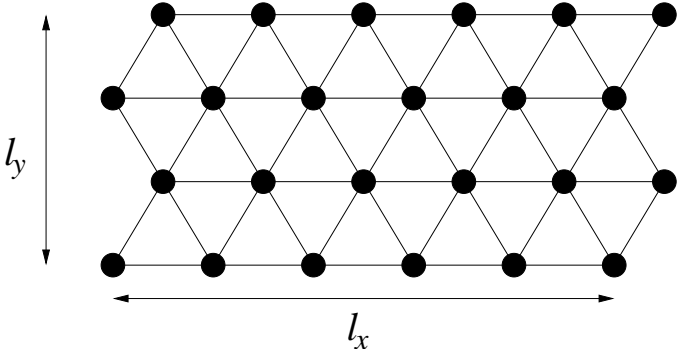}
   \caption{Illustration of the orientation of the triangular lattice.}
  \label{MULTI-triangularlattice}
\end{figure} 
In the antiferromagnetic regime, the exact
point of transition 
depends somewhat on the extent of the lattice in $x$-direction $l_x$
(see Fig.\ref{MULTI-triangularlattice}):
for example, it is $a_c=-1/3$ for $l_x = 3k$, $-0.353553$ for $l_x = 4k$,
$-0.350373$ for $l_x=5k$. For $l_x$ that are compound numbers, the
lowest absolute value of the transition for any of the divisors is the
relevant one. The complexity profile near the transition
(Fig. \ref{MULTI-triangularprof}) now shows slight
signs of frustration (small oscillations with $k$ and negative values for
$k\approx n$), but no large-amplitude oscillations as found for the
infinite range antiferromagnet. We interpret
this as a sign of localized frustrations, consistent with a large ensemble
of degenerate ground states, but with only local, rather than global, constraints.

\section{Summary}
\label{SEC-Results}
\begin{figure}
\epsfxsize= 0.8 \columnwidth
\epsffile{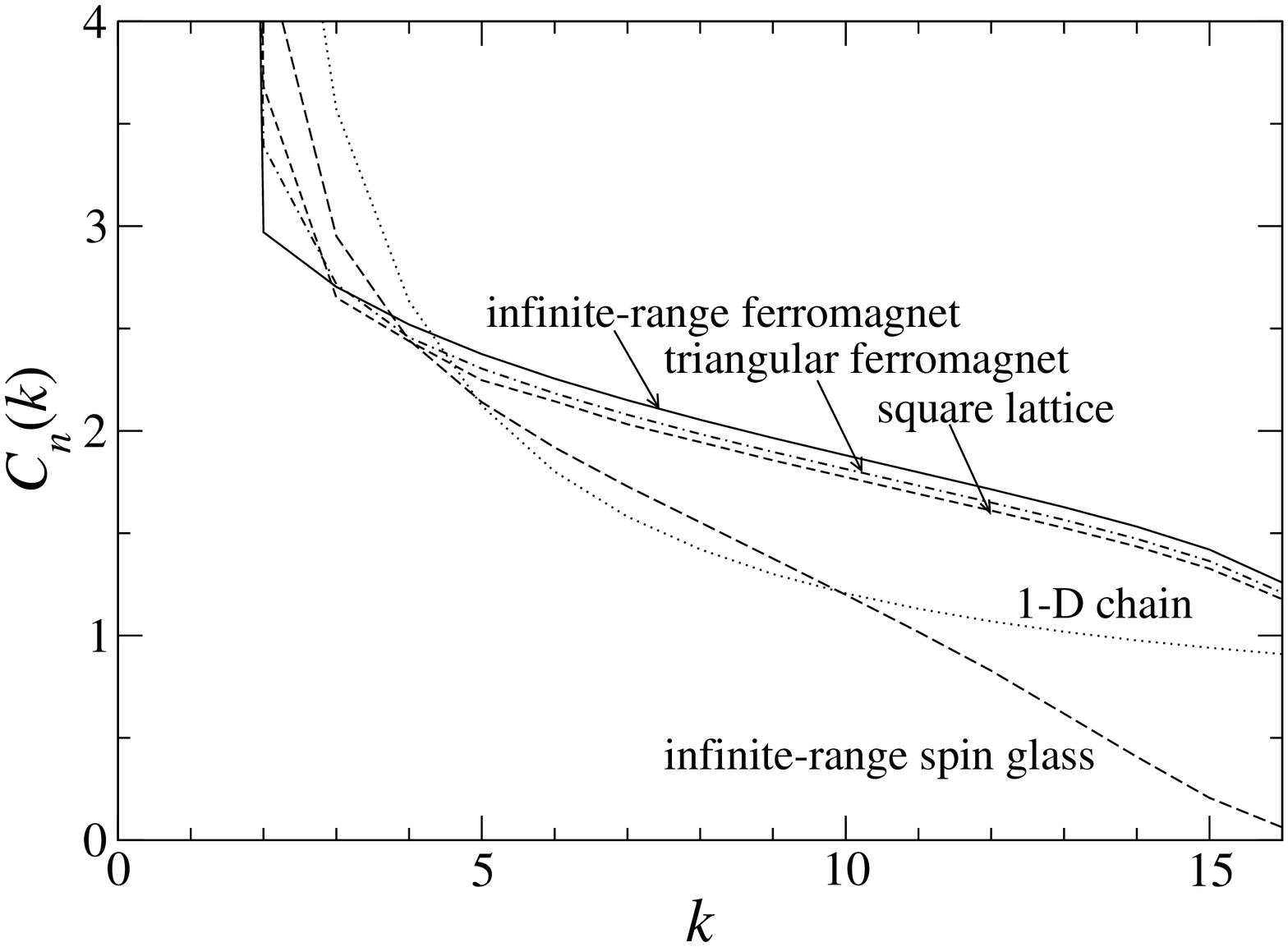}
   \caption{Complexity profiles for different interactions that show
     monotonically decreasing $C_n(k)$, at the same ratio $a/a_c = 0.999$,
     for $n=16$. The infinite-range ferromagnet, the square
lattice and the triangular lattice yield remarkably similar curves.}
  \label{MULTI-comparison}
\end{figure}

Comparing the multiscale complexity profiles that were found in the preceding
section, some intuitive and some rather surprising features emerge. 
The results are summarized in Table \ref{MULTI-summarytable} and
Fig. \ref{MULTI-comparison}.
All systems we considered
show a transition from well-behaved (finite magnitude) spins to 
diverging amplitude spins when the interaction between different spins
overrides the self-interaction. 

\begin{table}
\begin{tabular}{|l|l|l|l|}
\hline 
 {\bf Structure}& {\bf Transition: $a_c$ }& {\bf $C_n(k)$ extensive} & {\bf
   Oscillations} \\
\hline 
three-spin ferromagnet & $1/2$ &  N/A & no \\
three-spin antiferromagnet &$-1$ & N/A & yes \\   
inf.-range ferromagnet & $ 1/(n-1) $ &no & no \\
inf.-range antiferromagnet & $-1 $& no & yes \\
spin chain, $n$ even & $1/2$ & yes & no \\
square lattice, $L$ even & $1/4$ & yes & no \\
block matrices & varies & varies & varies \\
inf.-range spin glass & $\propto 1/\sqrt{n}$ & yes & no \\
triangular ferromagnet & $1/6$ & yes & no \\
triangular antiferromagnet & $\approx -1/3$& no & weak \\ \hline 
 \end{tabular}
\caption{Comparison of qualitative features from from Sections
  \ref{SEC-threespin} to \ref{SEC-triangular}.}
\label{MULTI-summarytable} 
\end{table}

We find two different universal behaviors of the complexity profile: 
monotonic decrease with scale, and oscillatory behavior. 
The former is a signature of systems with variables that are not frustrated.
At the transition, spins become increasingly redundant because
they are completely correlated (anticorrelated, in the case
of an antiferromagnet) with each other. 
Near the critical point, complexities diverge logarithmically. 
Fig. \ref{MULTI-comparison} shows a comparison
between the curves near divergence (at $0.999a_c$, for $n=16$) 
for the interaction structures that show this
behavior. The complexity profile can be interpreted as the cumulative
spectrum of collective behaviors of the system. 
The independence of $C(k,\rh)$ with changes in $\rh$ in all these cases implies 
that $D(k)=C(k+1)-C(k)$ is independent of $\rh$ near the transition. Thus, the
the collective behavior at $C(n)$ increases near the transition without
changing the spectrum of excitations at all scales between $1$ and $n$.
One can see that the infinite-range ferromagnet, the square lattice and the triangular lattice yield very similar curves, whereas the infinite-range spin glass and the 1-D chain show a quantitatively
different decay. For the simple ferromagnetic systems (all except the spin
glass) the value of $C_n(k)$ for large $k$ appears to increase monotonically 
with spin connectivity.

Away from the transition, in the weak-coupling regime, one finds that 
$C_n(k)$ decays exponentially with $k$.
This corresponds to fluctuations on
small length scales, as one expects from systems in the disordered
phase. In systems that have local interactions or disordered interactions
(essentially, in all the cases we studied except for the mean-field
ferromagnet and block matrices), multiscale complexities 
are approximately extensive, i.e., proportional to system size, for small
$k$ away from the transition. In the language of magnetic systems, this
means that one observes a finite size correlation length indicating
small patches of correlated spins, each of which give a contribution to complexity.
Finite-size effects (deviations from extensivity) 
are more pronounced for two-dimensional and
infinite-range models as compared to the 1-D spin chain, as is to be expected
in higher dimensional systems.

The complexities $C_n(k)$ for $k$ of order 1 thus represent localized
fluctuations, whereas nonvanishing $C_n(k)$ for $k/n$ of order $1$ represent
the emergence of collective behaviors on the scale of the system as a whole, and
whose complexity is therefore not extensive. Since the multiscale
complexity is a very general formalism, requiring only the joint
probability distribution for input, it can thus be used to describe
fluctuations and to identify phase transitions without explicitly 
choosing order parameters.

The second universal behavior, oscillations, is observed
when a global constraint leads to redundancy: in the simplest case, for the
long range antiferromagnet,
mutual repulsion enforces the constraint $\sum x=0$, implying any one spin 
is determined if all others are known.  We find that such constraints do not
result from random interactions found in spin-glasses, i.e. frustrated interactions
do not necessarily result in frustrated variables. There exist enough interactions that
are not frustrated to ensure collective behaviors of the ferromagnetic type. 
Oscillations do arise, however, for symmetric antiferromagnetic interactions, 
both between individual spins and between blocks of spins. 

Finally, we note that the characterization that we have provided using the
multiscale complexity is distinct from the usual characterization of the
correlations in spatial systems using a correlation length. The  
multiscale complexity does not require a spatial structure. It identifies the
aggregate size of fluctuations in terms of the number of participating
spins regardless of the topology of spatial or non-spatial interactions.
Thus it provides a more generally applicable characterization of the
collective behavior in interacting systems.

\section{Acknowledgments}
We thank Mehran Kardar for important suggestions and fruitful discussion.

\end{document}